\documentclass[aps,prb,showpacs,superscriptaddress,twocolumn,10pt,floatfix]{revtex4-1}
\usepackage[T1]{fontenc}
\usepackage{graphicx}
\usepackage{amsmath}
\usepackage{amssymb}
\usepackage{epsfig}
\usepackage{physics}
\usepackage{multirow}
\usepackage{siunitx}
\usepackage[dvipsnames]{xcolor}
\begin{document}

\title{Photocurrent and photoconductance of an helical edge state}

\author{Jonathan Atteia}
\email{jonathan.atteia@gmail.com}
\affiliation{Universit\'e de Bordeaux, CNRS, LOMA, UMR 5798, F-33405 Talence, France}

\author{J\'er\^ome Cayssol}
\email{jerome.cayssol@u-bordeaux.fr}
\affiliation{Universit\'e de Bordeaux, CNRS, LOMA, UMR 5798, F-33405 Talence, France}

\date{\today}
\begin{abstract}
We consider the helical edge state of a Quantum Spin Hall insulator, connected to leads, and irradiated by a monochromatic and circularly polarized electromagnetic wave. The photocurrent carried by a single helical edge state is studied as function of the characteristics of the electromagnetic radiation and electronic doping. We focus on the effect of the distance between the leads on the photocurrent. We also investigate the differential photoconductance of the helical edge state in presence of a small voltage bias between the leads.      
\end{abstract}

\maketitle

\section{Introduction}

Topological insulators (TI) are a novel class of materials that has attracted a lot of interest over the last decade due to their spectral and transport properties. TIs are insulating in the bulk and present robust, topologically protected metallic edge or surface states\cite{Hasan2010,Qi2011}. This protection is topological in the sense that a bulk gap closing is needed to suppress the conducting edge states. In two dimensions, the quantum spin Hall (QSH) effect is an example of a time-reversal invariant topological insulator that presents a helical liquid at its edge\cite{Kane2005a,Wu2006,Kane2005b}. The helical edge state consists in two counterpropagating edge channels where the direction of motion is locked to the value of the spin projection. The QSH effect was theoretically predicted and then observed experimentally in HgTe/CdTe quantum wells\cite{Bernevig2006,Konig2007}, later on in InAs/GaSb quantum wells\cite{Liu2008,Du2015}, and more recently in WTe$_2$ monolayers\cite{Qian2014,Wu2018}. However, the topological properties arise from the presence of spin-orbit interaction and of a band inversion process which requires a fine tuning of the band structure and makes it hard to engineer such TIs.

Recently, it was realized that driving periodically a topologically trivial material could allow to generate non-trivial phases of matter. Such phases were dubbed Floquet topological insulators\cite{Cayssol2013} due to the use of the Floquet theorem\cite{Shirley1965,Sambe1973}, the temporal analog of Bloch's theorem, to describe them. The frequency and strength of the external driving are thus additional parameters that allow to tune the band structure. For example, graphene\cite{Oka2009}, as well as a semi-conducting heterostructure\cite{Lindner2011} irradiated by a circularly polarized electromagnetic wave turn to a Chern insulator, a topological insulator where time-reversal symmetry is explicitely broken. Floquet edge states were thus predicted to appear at the boundaries of the material\cite{Kitagawa2011,Usaj2014,Perez-Piskunow2014,Gomez2014}. These discoveries led to the extension of the standard "periodic table" of equilibrium topological insulators\cite{Schnyder2008,Schnyder2009,Ryu2010} to an even richer classification of out-of-equilibrium Floquet topological insulators \cite{Kitagawa2010,Lindner2011,Gomez2013,Rudner2013,Carpentier2015}. In such systems, additional "sidebands", i.e. replicas of the original band dressed coherently by one or several photon, can contribute to the DC current, but also generate a non-equilibrium distribution in the irradiated region\cite{Kitagawa2011}. Several studies have shown that one can probe the peculiar Floquet spectrum and the edge state using electronic transport in a two- or multi-terminal setup\cite{Gu2011,Foa2014,Fruchart2016a,Atteia2017}. Experimentally, signatures of Floquet-Bloch states have been observed at the surface of a three-dimensional TI, Bi$_2$Se$_3$\cite{Wang2013}. The presence of edge states have been observed at the edge of a photonic Floquet topological insulator\cite{Rechtsman2013}. More recently, the anomalous Hall conductance was observed in graphene irradiated by a femtosecond laser pulse\cite{McIver2019}.

Besides, when a one-dimensional insulating chain is driven periodically in the adiabatic regime, an integer number of electrons is transferred through the chain during each cycle of the drive \cite{Thouless1983}. The number of transferred electrons corresponds to the Chern number of this Thouless pump. Such a quantized photocurrent was found to be present in the helical edge state of the QSH effect when the electrons are coupled to a rotating magnetic field through the Zeeman interaction\cite{Qi2008,Dora2012}. Dora \textit{et al.}\cite{Dora2012} found that upon increasing the frequency of the driving, a transition is triggered from a quantized photocurrent to a non-quantized transport regime. Subsequently, Vajna \textit{et al.}\cite{Vajna2016} extended this analysis of the photocurrent by introducing dissipation through coupling with different kinds of environments within a Lindblad master equation approach. More recently, it was pointed out that this charge transfer was originating from the chiral anomaly characteristic of Dirac fermions subject to external electromagnetic fields\cite{Fleckenstein2016}. It is also possible to photoexcite non-dispersive electron wavepackets in the helical edge state which was also shown to be a signature of the chiral anomaly\cite{Dolcini2016}.

In this work, we investigate the photocurrent and photoconductance carried through a single driven QSH helical edge state taking into account the finite length $L$ of the irradiated edge state between the two external leads (see Fig. \ref{fig:geometry}), while previous works focused on an infinite length helical state \cite{Dora2012,Vajna2016}. We use Landauer-B\"uttiker formalism extended to Floquet systems which allows us to compute the photocurrent of the irradiated edge state as a function of the frequency and strength of the driving field. In the low-frequency regime and when the chemical potential is at the band crossing, we recover the adiabatic quantized pumped current. However in the high-frequency regime, we find a different behavior than in the infinite edge limit \cite{Dora2012}, and we associate this discrepancy with the presence of leads. We provide an analytical formula that allows to describe the full crossover from $L=0$ (no irradiation and hence no photocurrent towards the $L \rightarrow \infty$ limit covered by previous works Refs \cite{Dora2012,Vajna2016}. The photocurrent is maximal at half-filling, and can be reduced by varying the chemical potential of the edge away from half-filling. Finally, we investigate the effect of the application of a potential bias and the corresponding differential photoconductance, which allows to scan the Floquet spectrum of the edge state.

This paper is organized as follows : in Sec. \ref{sec:model}, we describe the model and give a brief introduction to Floquet theory. In Sec. III, the Landauer-B\"uttiker formalism is used to compute the scattering coefficients of the finite length helical state. In Sec. \ref{sec:results}, we present the results for the photocurrent and the photoconductance. Conclusions are given in Sec. \ref{sec:conclusion}.

\section{Model and Floquet Hamiltonian}

\label{sec:model}

In this section, we introduce the Hamiltonian of the helical edge state of the quantum spin Hall insulator (QSH) when it is irradiated by a monochromatic and circularly polarized electromagnetic wave. We also review the quasi-stationnary Floquet states of the QSH helical edge under irradiation.

\subsection{Model}

\begin{figure}[h!]
	\includegraphics[width=7cm]{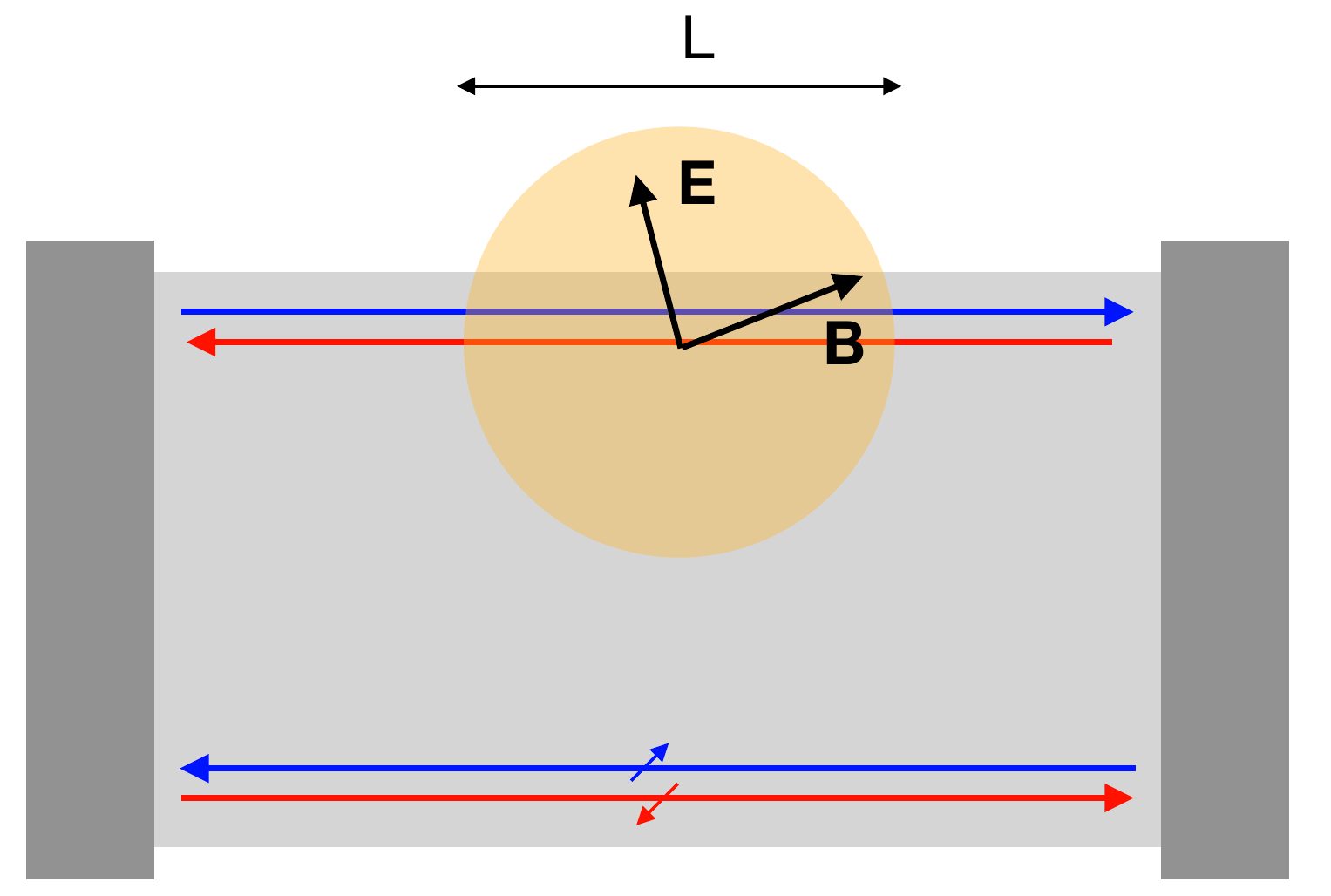} \\
	\caption{A quantum spin Hall insulator is connected to two reservoir leads and the helical edge state is irradiated by a circularly polarized electromagnetic wave over a region of length $L$. Spin up edge channel is represented in blue and rotate clockwise while spin down electrons (red arrows) rotate counter-clockwise. The yellow shaded area represents the laser spot. The focused laser spot and the two-terminal set-up allow to extract the pumped current carried by a single edge state (here the upper edge) by using a low impedance external circuit to avoid reinjection of the photocurrent towards the lower edge}.
	\label{fig:geometry}
\end{figure}

We consider a QSH insulator (located in the $xy$ plane) connected to two leads. The helical edge state is irradiated over a region of length $L$ which is also the length of the edge state between the two leads [Fig. \ref{fig:geometry}]. 
 
The QSH edge state couples with the electromagnetic wave to charge through the vector potential $\mathbf{A}=\mathbf{E}/\omega=A_0(\cos(\omega t),\sin(\omega t))$ and to the spin through Zeeman interaction with the magnetic field $\mathbf{B}(t)=-B_0(\cos\omega t,\sin\omega t)$ with $B_0=\omega/cA_0$. The Hamiltonian of the irradiated region reads :
\begin{equation}
H(k,t)=vk\sigma_z-\mu+evA_0\sigma_z\cos(\omega t)+g[\sigma_+e^{-i\omega t}+\sigma_-e^{i\omega t}],
\label{eq:Hamiltonian}
\end{equation}
where $k$ is the one-dimensional wavevector of an electron along the edge channel, $\mu$ the chemical potential, and $\sigma_\pm=(\sigma_x\pm i\sigma_y)/2$ are the raising/lowering spin operators. The matrices $\sigma_x,\sigma_y,\sigma_z$ are the standard spin Pauli matrices. The Zeeman coupling is characterized by $g=\mu_Bg_\text{eff}B_0$, where $\mu_B$ is the Bohr magneton and $g_\text{eff}$ the effective Land\'e factor. We have set the Fermi velocity to one and $\hbar\equiv1$. Due to the Zeeman coupling, the rotating magnetic field allows transitions between the two spin projections, which are also transitions between right and left movers. Moreover, $\hbar\omega$ must be smaller than the bulk band gap of the QSH insulator. For a band gap of $100$ meV \cite{Wu2018}, this restricts the frequencies to lower than $20$ THz. 

The effect of the orbital coupling can be neglected compared to the Zeeman term in the limit $evA_0\ll\omega$. We present here a simple argument to picture this, while we prove it in the next section. Indeed, the orbital coupling strength is $evA_0$, with $A_0=E_0/\omega$, where $E_0$ the electric field strength. Although the interaction strength of the orbital part is more important than the Zeeman strength ($evA_0\gg g$ for frequencies of the order of the THz), it cannot generate transitions between the eigenstates of $\sigma_z$, and the energy originating from the coupling to the electric field becomes thus negligible compared to the quantum of energy $\hbar\omega$ that can be absorbed, namely $evE_0/\omega\ll\hbar\omega$.

\subsection{Floquet spectrum and eigenstates}

The quasi-eigenstates and quasi-energies of the system are obtained using the Floquet theory for periodically driven systems. First we consider the spectrum of the infinite system and fix the momentum $k$. Due to the time-periodicity of the driving $H(k,t)=H(k,t+T)$, we use the Floquet theorem to solve the time-dependent Schr\"odinger equation under the form :
\begin{equation}
\Psi(k,t)=e^{-i\varepsilon t/\hbar}\Phi(k,t)  \, ,
\label{eq:Floquet_theo}
\end{equation}
where $\varepsilon$ is the quasi-energy which is defined modulo $\hbar\omega$ and $\Phi(k,t)=\Phi(k,t+T)$ is a time-periodic function, associated to the quasienergy $\epsilon$. Injecting Eq. (\ref{eq:Floquet_theo}) into the Schr\"odinger equation provides the following eigenvalue equation for $\Phi(k,t)$ and $\epsilon$ :
\begin{equation}
(H(k,t)-i\partial_t)\Phi(k,t)=\varepsilon\Phi(k,t)  \, .
\label{eq:FloquetHam}
\end{equation}
The effect of the vector potential can be captured through a unitary transformation $\Phi(k,t)=U(t)\tilde{\Phi}(k,t)$ with $U(t)=\exp({i\sigma_zevA_0\sin(\omega t)/\omega})$ such that the Hamiltonian $\tilde{H}(k,t)=U^\dagger(t)(H(k,t)-i\partial_t)U(t)$ becomes \cite{Dora2012}:
\begin{align}
    \tilde{H}(k,t)&=vk\sigma_z+g\left[\sigma_+\exp(-i\omega t-i2evA_0\sin(\omega t)/\omega)\right.\nonumber \\
    &+\left. \sigma_-\exp(i\omega t+i2evA_0\sin(\omega t)/\omega)\right].
\end{align}
The exponential terms containing the vector potential can be expanded using the Jacobi-Anger formula $\exp(i2z\sin(\omega t))=\sum_{m=-\infty}^\infty J_m(2z)\exp(im\omega t)$ where $J_m(z)$ is the Bessel function of first order. In the limit $z=evA_0/\omega\ll1$ considered here, the Bessel function can be expanded as $J_m(2z)\rightarrow (\text{sign}(m)z)^{|m|}/|m|!$. In the rotated frame, the effective coupling between the Floquet replicas is thus proportional to $g(evA_0/\omega)^{|m|}$ and one can thus consider only the $m=0$ component. In the following of the paper, we consider only the limit $evA_0/\omega\ll1$, wherein the orbital part of electromagnetic coupling can be safely neglected.

The Hamiltonian (\ref{eq:FloquetHam}) can be solved in frequency space by expanding $\Phi(k,t)$ and $H(k,t)$ in Fourier series :
\begin{eqnarray}
    H(k,t)&=\sum_{m\in\mathbb{Z}}H_m(k)e^{-im\omega t} \, , \\
    \Phi(k,t)&=\sum_{m\in\mathbb{Z}}\Phi_{m}(k)e^{-im\omega t}  \, .
\end{eqnarray}
Inserting these expressions in Eq. (\ref{eq:FloquetHam}) leads to an infinite matrix equation in Floquet-Fourier space :
\begin{equation}
    \sum_{n\in\mathbb{Z}}(H_{m-n}(k)+n\hbar\omega\delta_{mn})\Phi_{n}(k)=\varepsilon\Phi_{m}(k) \, .
    \label{eq:FloquetFourier}
\end{equation}
In the specific case of the helical edge state, the Hamiltonian Eq.(\ref{eq:Hamiltonian}) contains only the zero and first order Fourier harmonics: $m=0,\pm 1$, while all the higher harmonics vanish, $H_m =0$ for all $|m| > 1$. Hence, for the irradiated helical liquid, the Eq. (\ref{eq:FloquetFourier}) reads :
\begin{equation}
    g\sigma_+\Phi_{n-1}+( k \sigma_z +n\omega)\Phi_{n}+g\sigma_-\Phi_{n+1}=\varepsilon\Phi_{n}  \, .
    \label{eq:recurrence2}
\end{equation}
Introducing the notation $\Phi_n=(u_n,v_n) ^T$, the components of the spinor $\Phi_n$ are shown to obey the infinite system of linear equations :
\begin{subequations}
\begin{align}
&gv_{n-1}+\left(k+n\omega\right)u_n=\varepsilon u_n \, , \\
&\left(-k+n\omega\right)v_{n}+gu_{n+1}=\varepsilon v_{n} \, .
\end{align}
\label{eq:diff}
\end{subequations}
Since each $u_n$ is only coupled to $v_{n-1}$, we can block-diagonalize the infinite system of equations by re-arranging the components into blocks $(u_n,v_{n-1}) ^T$. For each block $(u_n,v_{n-1}) ^T$, two quasi-energies are obtained for each value of the momentum $k$ :
\begin{equation}
\varepsilon_{\alpha,n}=n\omega-\frac{\omega}{2}+\alpha \sqrt{(k-\omega/2)^2+g^2} \, ,
\label{eq:disp}
\end{equation}
labelled by $\alpha=\pm 1$, and their associated eigenspinors $(u_n,v_{n-1}) ^T$ read :
\begin{equation}
\begin{pmatrix}
u_{\alpha,n} \\ 
v_{\alpha,n-1}
\end{pmatrix}=\frac{1}{\sqrt{2\lambda}}
\begin{pmatrix}
\sqrt{\lambda+\alpha(k-\omega/2)} \\ \alpha\sqrt{\lambda-\alpha(k-\omega/2)}
\end{pmatrix} \, ,
\label{eq:eig_irrad}
\end{equation}
where $\lambda=\sqrt{(k-\omega/2)^2+g^2}$ is half of the gap between the two quasi-energies corresponding to a fixed $n$ at a given momentum $k$. 

\begin{figure}[t]
	\includegraphics[width=8cm]{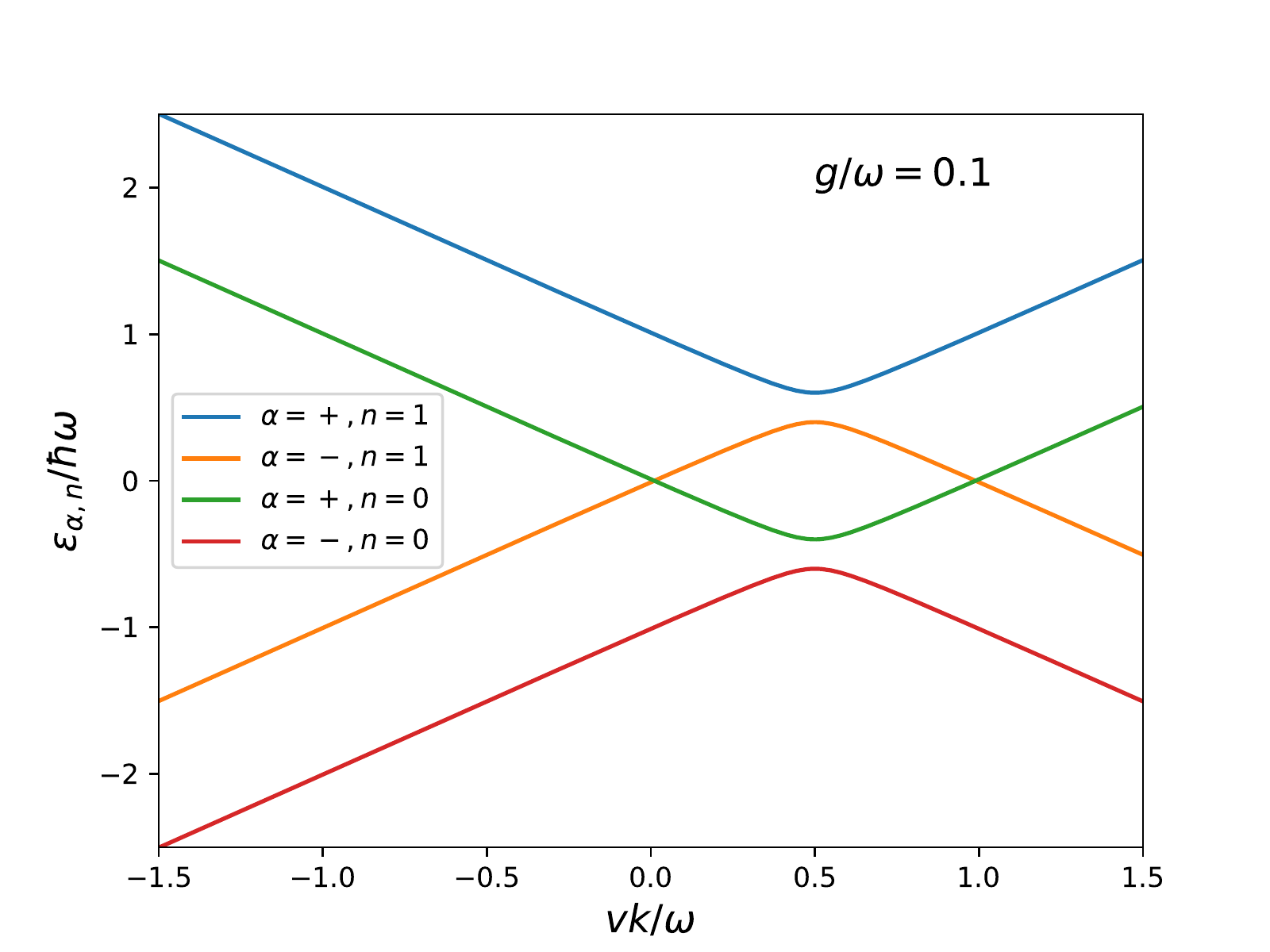} \\
	\includegraphics[width=8cm]{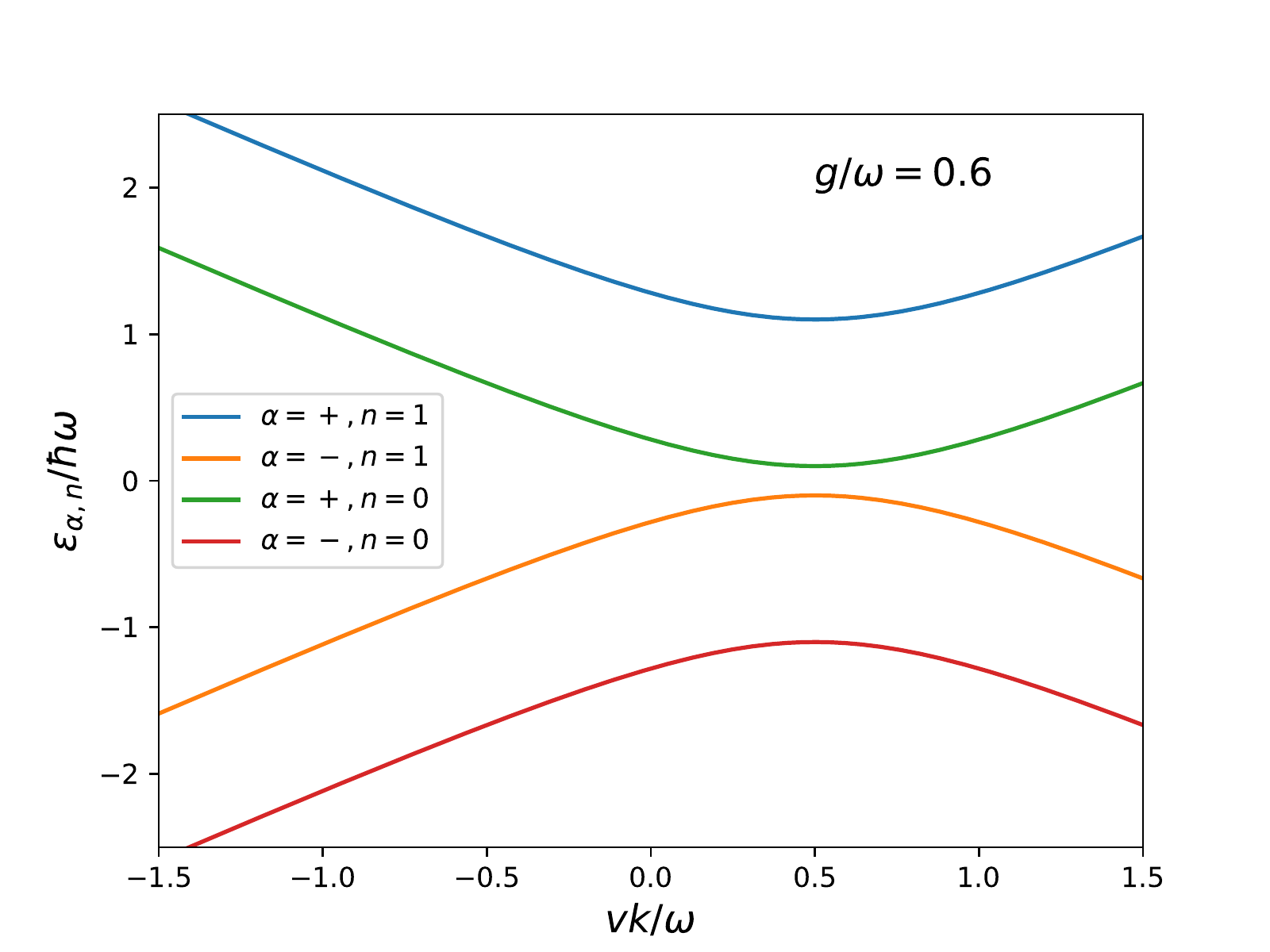} \\
	\caption{Quasi-energy spectra of the irradiated edge state for a driving strength a) $g=0.1\hbar\omega$ and b) $g=0.6\hbar\omega$. Only the Floquet replicas $n=0$ and $n=1$ are represented. We observe a gap in the spectra of size $2g$ located at $\varepsilon=\hbar\omega/2$ for the replica $n=1$ and at $\varepsilon=-\hbar\omega/2$ for the replica $n=0$. }
	\label{fig:spectrum}
\end{figure}

Fig. \ref{fig:spectrum} shows the dispersion relation of the $n=1$ and $n=0$ Floquet replicas. A gap of size $2g$ opens in the dispersion relation at momentum $k=\omega/2$ and quasi-energy $\varepsilon=n\omega+\omega/2$. For weak driving ($g<\omega/2$), this gap is located at $\varepsilon=\omega/2$ on the $n=1$ replica, while it is located at $\varepsilon=-\omega/2$, on the $n=0$ replica. This gap is thus located at the edges of the first Floquet zone. As the driving increases (frequency decreases), the band $(\alpha,n)=(+,0)$ rises above the band $(-,1)$ and a gap of size $2g$ opens at the Floquet zone center $\varepsilon=0$. This gap opening happens for $g=\omega/2$. Of course, there is an infinite set of replicas that are not represented on Fig. \ref{fig:spectrum}, but we plot only the replicas $n=0$ and $n=1$ which are the only replicas relevant for transport as we will see later.

\section{Scattering problem}

In this section, the scattering problem is formulated and solved within the Landauer-B\"{u}ttiker formalism extended to driven Floquet systems \cite{Moskalets2002}. We consider an helical edge state irradiated along a region of length $L$ and connected to (non irradiated) leads at $x=0$ and $x=L$. Due to the coupling with light, electrons injected at a given energy from one lead can be transmitted or reflected within different sidebands in the leads. We first write the wavefunctions in the various regions: the irradiated region ($0<x<L$) and the electrodes at left ($x<0$) and right ($x>0$). Because time-reversal invariance is broken by the presence of the driving, the transmission of electrons from the left lead to the right lead is not necessarily identical to the reverse process. We first calculate the reflection coefficient $R(\varepsilon)$ and transmission coefficient $T(\varepsilon)$ from left to right. Finally the coefficients from the reverse scattering problem, $R'(\varepsilon)$ and $T'(\varepsilon)$, are also provided at the end of this section.   

\subsection{Irradiated region}
In order to find the set of eigenstates at quasi-energy $\varepsilon$, we invert the dispersion relation (\ref{eq:disp}) :
\begin{equation}
k_{\alpha,n}=\frac{\omega}{2}+\alpha K_n,
\label{eq:disp_inv}
\end{equation}
where : 
\begin{equation}
K_n=\sqrt{(\varepsilon-n\omega+\omega/2)^2-g^2} \, .
\label{eq:disp_inv2}
\end{equation}
The quantity $K_n$ can be either real or purely imaginary depending on the sign of the expression below the radical. As a consequence, if the quasi-energy $\varepsilon$ is located within a gap, the wavevector $k_{\alpha,n}$ will be complex, which corresponds to evanescent states. The eigenstates at the quasi-energy $\varepsilon$ corresponding to the quasi-momentum $k_{\alpha,n}$ are written for each block $(u_n,v_{n-1}) ^T$ as :
\begin{equation}
\Tilde{\Phi}_{\alpha,n}(x,t)=
\begin{pmatrix}
u_{\alpha,n}e^{-in\omega t} \\ 
v_{\alpha,n-1}e^{-i(n-1)\omega t}
\end{pmatrix}
e^{ik_{\alpha,n}x} \, ,
\label{eq:eig_eps}
\end{equation}
with : 
\begin{align}
u_{\alpha,n}&=\frac{g}{\sqrt{g^2+(\varepsilon-n\omega-\omega/2-\alpha K_n)^2}}, \label{eq:spi_u} \\
v_{\alpha,n-1}&=\frac{(\varepsilon-n\omega-\omega/2-\alpha K_n)}{\sqrt{g^2+(\varepsilon-n\omega-\omega/2-\alpha K_n)^2}} \, . \label{eq:spi_v}
\end{align}
Summing all Fourier harmonics labelled by $n$, the full time-dependent wavefunction in the irradiated region can be written as $\Psi_I(x,t)=e^{-i\varepsilon t} \Phi_I(x,t)$ with time-periodic function given by :
\begin{align}
\Phi_I	&=\sum_n\left( a_n \Tilde{\Phi}_{+,n}(x,t)+b_n \Tilde{\Phi}_{-,n}(x,t)\right) \\
	&=\sum_n
	\begin{pmatrix}a_nu_{+,n}e^{ik_{+,n}x}+b_nu_{-,n}e^{ik_{-,n}x}\\
	a_{n+1}v_{+,n}e^{ik_{+,n+1}x}+b_{n+1}v_{-,n}e^{ik_{-,n+1}x}
	\end{pmatrix}
	e^{-in\omega t} \, ,
\end{align}
where $a_n$ and $b_n$ are complex amplitudes to be determined by imposing the matching conditions at the interfaces $x=0$ and $x=L$ with the leads. In the last line, the components of the spinors have been rearranged Fourier harmonics by Fourier harmonics in view of writing down the continuity of the wavefunction (see section C. below).

\subsection{Wavefunctions in the leads}

In order to obtain the transmission and reflection coefficients, we construct the wavefunction in the leads. We first calculate the transmission coefficients from left to right, namely with an electron of energy $\varepsilon$ incoming from the left electrode. 

{\it Left lead (L):} In the leads, the wavefunction is made of an electron incoming at energy $\varepsilon$ and the sum of all the reflected electrons at energy $\varepsilon+n\hbar\omega$ :
\begin{equation}
\Psi_L(x,t)=e^{-i\varepsilon t}\left(\phi^+e^{ik_nx}+\sum_nr_n\phi^-e^{-ik_nx}e^{-in\omega t}\right),
\end{equation}
where $k_n=\varepsilon-n\omega$, $\sigma_z\phi^\pm=\pm\phi^\pm$ are the left/right $(+/-)$ movers and $r_n$ are the reflection coefficients for electrons exiting with energy $\varepsilon+n\omega$. 

{\it Right lead (R):} The wavefunction in the right lead is expressed as :
\begin{equation}
\Psi_R(x,t)=e^{-i\varepsilon t}\sum_nt_n\phi^+e^{ik_n(x-L)}e^{-in\omega t},
\end{equation}
where $t_n$ denote the transmission coefficients towards the various sidebands (indexed by $n$) of the right lead. 

\subsection{Matching of the wavefunctions}

The wavefunctions in the various regions are matched together by writing the continuity of the spinors at the interfaces at any time $t$, which reads : $\Psi_L(x=0,t)=\Psi_I(x=0,t)$ and $\Psi_I(x=L,t)=\Psi_R(x=L,t)$. This is equivalent to matching all the Fourier components of these spinor wavefunctions, which yields to the following system of equations :
\begin{align}
&a_nu_{+,n}+b_nu_{-,n}=\delta_{n0}, \label{eq:sys1} \\
&a_{n+1}v_{+,n}+b_{n+1}v_{-,n}=r_n, \label{eq:sys2} \\
&a_nu_{+,n}e^{ik_{+,n}L}+b_nu_{-,n}e^{ik_{-,n}L}=t_n, \label{eq:sys3} \\
&a_{n+1}v_{+,n}e^{ik_{+,n+1}L}+b_{n+1}v_{-,n}e^{ik_{-,n+1}L}=0. \label{eq:sys4}
\end{align}
In the case $n\neq0$, Eqs. (\ref{eq:sys1}) and (\ref{eq:sys4}) form a homogeneous linear system of equations which has no solution apart from the trivial one. Thus, we must have $a_n=b_n=0$ for any non zero  $n\neq0$. We can deduce from Eqs. (\ref{eq:sys2}) and (\ref{eq:sys3}) that the only non-vanishing coefficients are $r_{-1}$ and $t_0$. This system becomes :
\begin{align}
&au_++bu_-=1, \label{eq:sysB1} \\
&av_++bv_-=r_{-1}, \label{eq:sysB2} \\
&au_+e^{ik_{+,0}L}+bu_-e^{ik_{-,0}L}=t_0, \label{eq:sysB3} \\
&av_+e^{ik_{+,0}L}+bv_-e^{ik_{-,0}L}=0. \label{eq:sysB4}
\end{align}
where we have set $a\equiv a_0$, $b\equiv b_0$, $u_\pm=u_{\pm,0}$ and $v_\pm=v_{\pm,-1}$. We can see that an electron can only be transmitted at the same energy, or be reflected having emitted one photon. For the case of an electron originating from the left lead, only the spinor $(u_0,v_{-1})$ contributes to the scattering process.


\subsection{Reflection and transmission coefficients : incoming electron from the left lead}

Solving the linear system Eqs. (\ref{eq:sysB1}) and (\ref{eq:sysB4}) for $a$ and $b$, and then using Eqs. (\ref{eq:sysB2}) and (\ref{eq:sysB3}) to evaluate the amplitudes $r_{-1}$ and $t_0$, yields the transmission and reflection probabilities :
\begin{align}
T(\varepsilon)=|t_0|^2=\frac{K_0^2}{K_0^2+g^2\sin^2(K_0L)}, \label{eq:transLR} \\
R(\varepsilon)=|r_{-1}|^2=\frac{g^2\sin^2(K_0L)}{K_0^2+g^2\sin^2(K_0L)} \, ,
\end{align}
where $K_0=\sqrt{(\varepsilon+\omega/2)^2-g^2}$, and where expressions (\ref{eq:spi_u}) and (\ref{eq:spi_v}) have been used. When the energy is in the gap, $(\varepsilon+\omega/2)^2<g^2$, $K_0$ is imaginary and the current is carried by evanescent states.

\begin{figure}[t]
	\includegraphics[width=8cm]{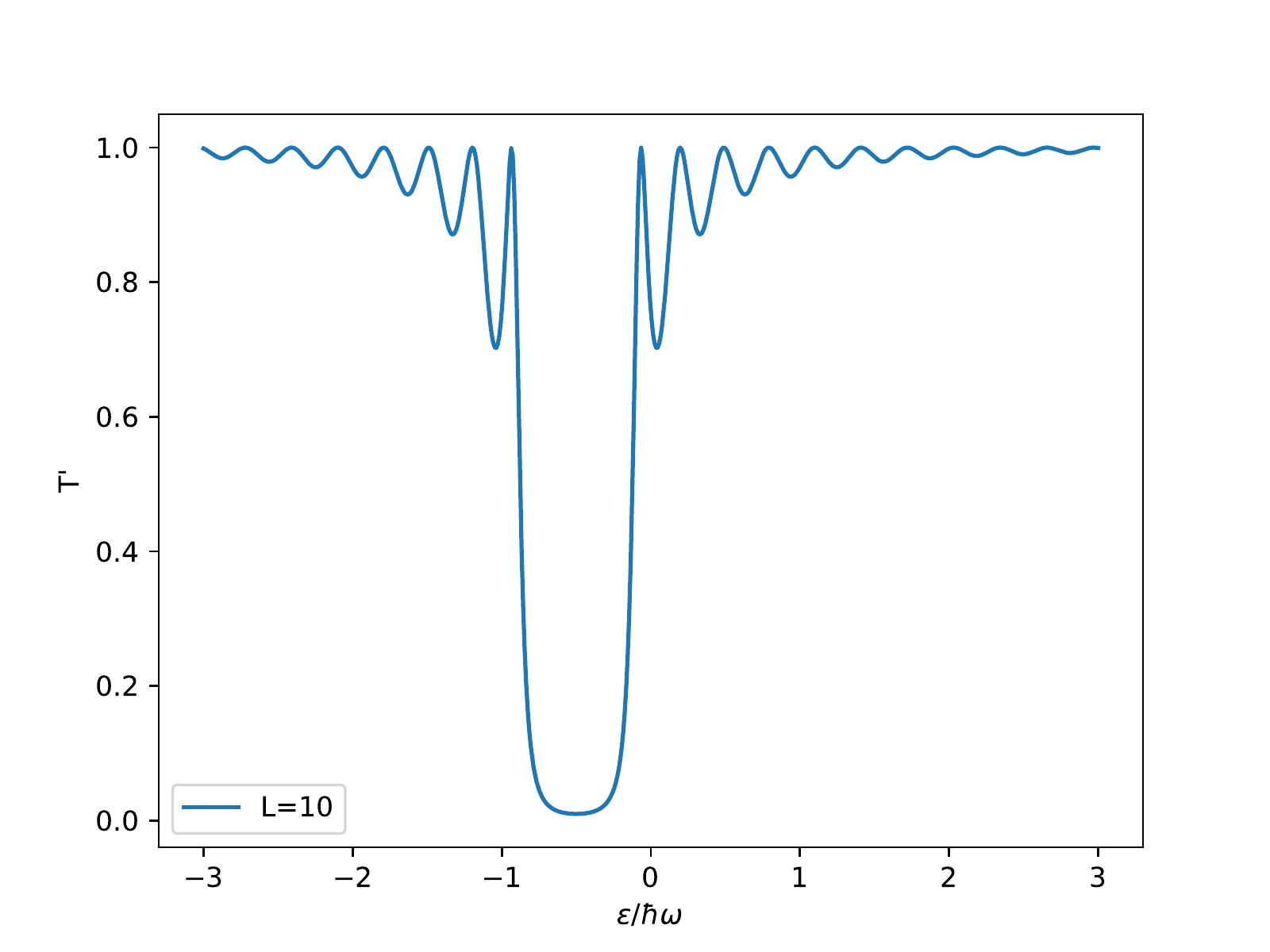} \\
	\caption{Transmission probability $T(\varepsilon)$ given by Eq. (\ref{eq:transLR} for an electron originating from the left lead to exit in the right lead as a function of the energy of the incident electron for a driving strength $g=0.3\hbar\omega$ and length $L=10\hbar v/\omega$. We observe a dip in the transmission probability centered at $\varepsilon=-\omega/2$ and of width $2g$ which corresponds to a reflected electron having emitted one photon. The oscillations corresponds to Fabry-P\'erot interferences and are thus dependent on the length of the sample.}
	\label{fig:trans_coeffs}
\end{figure}

Fig. \ref{fig:trans_coeffs} shows the transmission probability as a function of the incoming electron energy. We observe a dip in the transmission probability at $\varepsilon=-\omega/2$ which corresponds to the one photon resonance between the conduction band and the valence band. For a short ribbon, the transmission probability in the gap is reduced but doesn't vanish, which means that the current is carried by evanescent states. As the length of the ribbon is increased, the dip gets sharper and no evanescent states contribute to the current. The oscillations correspond to Fabry-P\'erot interferences.
\subsection{Reflection and transmission coefficients : incoming electron from the right lead}
Applying the same reasoning for an electron incoming from the right lead and exiting in the left lead, we find the transmission and reflection coefficients :
\begin{align}
T'(\varepsilon)=|t'_0|^2=\frac{K_1^2}{K_1^2+g^2\sin^2(K_1L)}, \label{eq:transRL} \\
R'(\varepsilon)=|r'_{1}|^2=\frac{g^2\sin^2(K_1L)}{K_1^2+g^2\sin^2(K_1L)},
\end{align}
with $K_1=\sqrt{(\varepsilon-\omega/2)^2-g^2}$. The probabilities are identical to the one for electrons incoming from the left lead except for the shift $\omega$ in energy.

\section{Photocurrent and photoconductance}

\label{sec:results}

In this section, we present our results for the photocurrent and the photoconductance of a finite length helical edge state contacted by two electrodes, using the scattering amplitudes computed in the section III. In the absence of any voltage bias between the electrodes, a finite photocurrent flows along the edge. We also compute the differential photoconductance as function of a voltage bias between the leads.

\subsection{Pumped photocurrent}

\label{sec:pumped_current}

Because time-reversal symmetry is broken (due to the circularly polarized light), the transmission probabilities from electrons incoming from the left and right leads are not identical. Such an asymmetry allows for the generation of a pumped current in the absence of a potential difference between the leads. 

\subsubsection{Formula for the pumped current at zero temperature}

According to the Landauer-B\"uttiker formalism extended to Floquet systems, the DC current through the irradiated edge state is equal to \cite{Moskalets2002,Atteia2017} :
\begin{align}
I(\mu)&=\frac{e}{h}\int_{-\Lambda}^{\Lambda}d\varepsilon \sum_n \left(T_n(\varepsilon) f_L(\varepsilon) - T'_n(\varepsilon) f_R(\varepsilon) \right), 
\label{eq:current}
\end{align}
where $T_n(\varepsilon)$ is the probability of an electron incoming from the left lead to be transmitted in the right lead having absorbed $n$ photons, and $T'_n(\varepsilon)$ is the reversed process. The functions $f_L(\varepsilon)$ and $f_R(\varepsilon)$ are the Fermi distributions at chemical potential $\mu$ in the left and right leads respectively. The cut-off $\Lambda$ in the integral that represents the bandwidth of the edge state, namely the bulk gap of the material. Because the electrons can only be transmitted in the channel $n=0$, we have $T_n(\varepsilon)=T(\varepsilon)$ and $T'_n(\varepsilon)= T'(\varepsilon)$. At zero bias and zero temperature, the photo-current simply reads :
\begin{align}
I(\mu)=\frac{e}{h}\int_{-\Lambda}^{\mu}d\varepsilon \left(T(\varepsilon)- T'(\varepsilon) \right).
\label{eq:pumped_diff}
\end{align}
The pumped current is therefore the difference between transmission from the left to the right lead and the reverse process. The function $T(\varepsilon)$ and $T'(\varepsilon)$ are identical except for the shift $\pm\hbar\omega/2$ in energy. In order to simplify the expression for the current, we make a change of variable $\pm\hbar \omega/2$ in each integral, and define the symmetric function :
\begin{equation}
P(x)=\frac{x^2-g^2}{x^2-g^2+g^2\sin(\sqrt{x^2-g^2}L)} \, ,
\label{P}
\end{equation}
such that :
\begin{align}
    T(\varepsilon)&=P(\varepsilon+\omega/2), \\
    T'(\varepsilon)&=P(\varepsilon-\omega/2).
\end{align}
The function $P(\varepsilon)$ is identical to the function $T(\varepsilon)$ plotted on Fig. \ref{fig:trans_coeffs} except that the dip of width $2g$ is centered on $\varepsilon=0$ instead of $\varepsilon=-\hbar\omega/2$. For a long ribbon, the dip is well defined and corresponds to the gap of size $2g$ in the dispersion relation. The current has the expression :
\begin{subequations}
\begin{align}
I(\mu)&=\frac{e}{h}\int_{-\Lambda-\omega/2}^{\mu-\omega/2}d\varepsilon P(\varepsilon)- \frac{e}{h}\int_{-\Lambda+\omega/2}^{\mu+\omega/2}d\varepsilon P(\varepsilon) \\
&=\frac{e}{h}\int_{-\Lambda-\omega/2}^{-\Lambda+\omega/2}d\varepsilon P(\varepsilon)- \frac{e}{h}\int_{\mu-\omega/2}^{\mu+\omega/2}d\varepsilon P(\varepsilon).
\end{align}
\label{eq:derivation_curr}
\end{subequations}

We can see that in the limit $\Lambda\rightarrow\infty$, we have $P(\varepsilon\rightarrow-\infty)=1$, thus :
\begin{align}
I(\mu)&=\frac{e\omega}{2\pi}- \frac{e}{h}\int_{\mu-\omega/2}^{\mu+ \omega/2}d\varepsilon P(\varepsilon).
\label{eq:pumped_current}
\end{align}

The pumped current is the sum of two terms. The first term corresponds to a quantized pumped charge per unit cycle such that $\int_0^T Idt=e$. We can see from Eqs. (\ref{eq:derivation_curr}) that this term originates from states located deep in the band of the edge state. This term has a topological origin \cite{Cayssol2013} which is related to Thouless's charge pumping mechanism \cite{Thouless1983}. For a left circular polarization, this current is directed along the positive $x$ axis. As the frequency is increased such that $\hbar\omega/2>g$, the second term generates a pumped current in the opposite direction. This current is carried by propagative states close the Fermi level such that incoming states can absorb or emit a quantum $\hbar\omega$. 

Finally, we have two ways of interpreting the expressions for the current. One way is to use the non-symmetrized expressions for the transmission probabilities (\ref{eq:transLR}) and (\ref{eq:transRL}) and consider the net transmission probabilities when the quasi-energy is in one of the gaps. This allows for a clear microscopic interpretation. The other way consists in using the symmetrized expression (\ref{eq:pumped_current}) which is simpler to compute the current (because the cut-off $\Lambda$ has disappeared from the equations) and allows to discriminate the two (high- and low-) frequency regimes. In order to interpret the results, we will switch back and forth from these two pictures.

\subsubsection{Length dependence of the pumped current} 
\begin{figure}[h]
	\begin{center}
		\includegraphics[width=8cm]{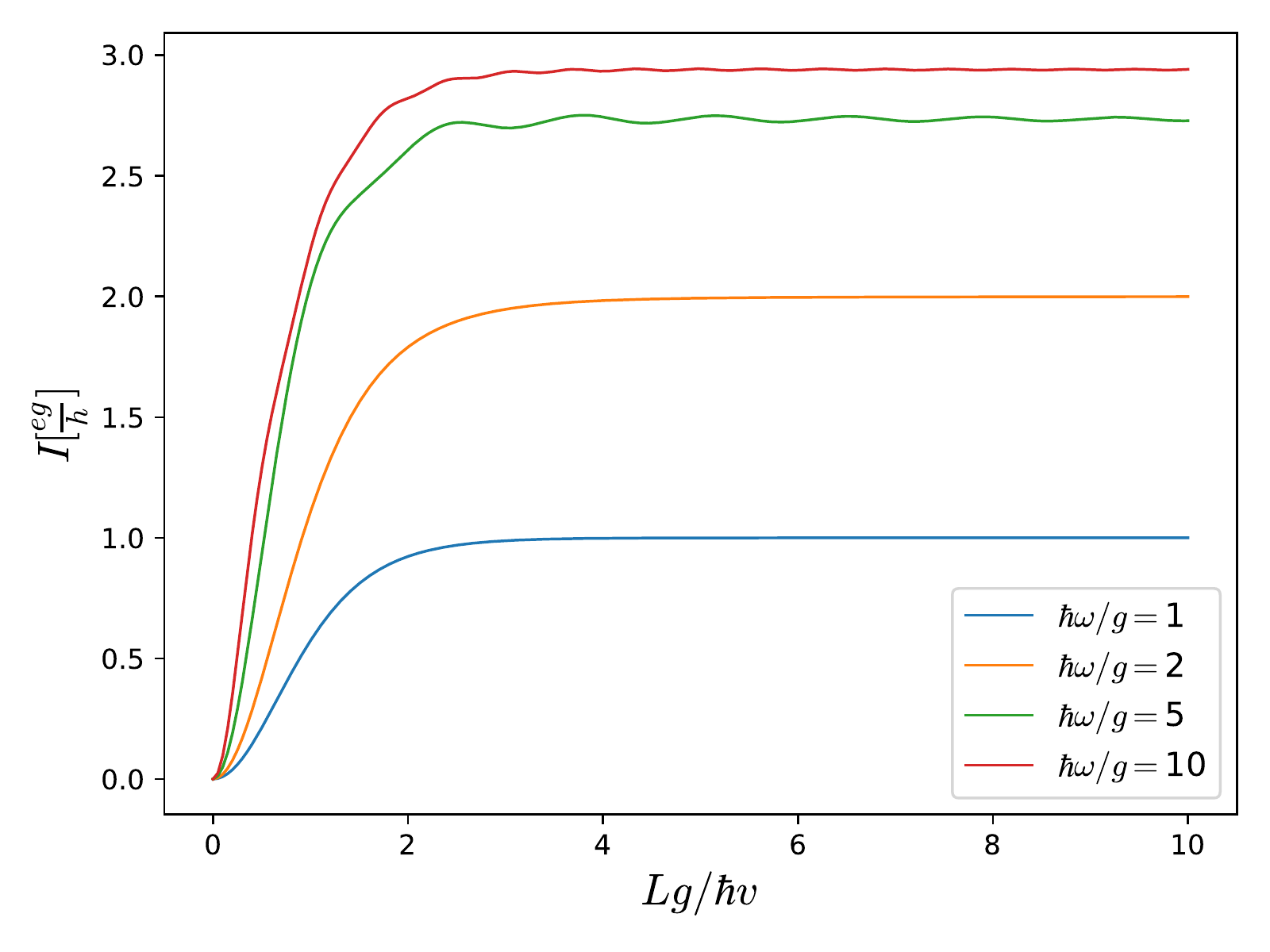}
		\caption{Pumped current as a function of the length $L$ of the irradiated region for different frequency regimes. The pumped current vanishes for $L=0$ and reaches a saturation regime for $L\gtrsim 2g/\hbar v$. This saturation is the $L$-independent value $e \omega/(2 \pi)$ at low frequency (two lower curves). At higher frequencies (two upper curves), the saturated current exhibits tiny Fabry-P\'erot like oscillations. The maximal photocurrent is $\pi eg/h$ and is reached in the limit of high frequency $\hbar \omega \gg g$ and long irradiated length $L \gg \xi$.}
		\label{fig:pumped_L}
	\end{center}
\end{figure}

We consider the pumped current at half-filling, namely $\mu=0$. The pumped current reads :
\begin{align}
I_{\mu=0}=\frac{e\omega}{2\pi}-2\frac{e}{h} \int_{0}^{\hbar\omega/2}d\varepsilon \, P(\varepsilon), \label{eq:pumped}
\end{align}
where $P(\epsilon)$ is given by Eq. (\ref{P}). The analytical expressions Eqs. (\ref{P},\ref{eq:pumped}) allow to study how the pumped current depends upon the length of the irradiated region (Fig. \ref{fig:pumped_L}). In the limit $L \rightarrow 0$, the pumped current obviously vanishes (no irradiation). In the opposite limit $L \rightarrow \infty$ (corresponding to Refs \cite{Dora2012,Vajna2016}), the pumped current reaches a saturation regime which depends on the ratio $\omega/g$ (Fig. \ref{fig:pumped_L}). The crossover between these two limits arises about the characteristic length $\xi=1/\kappa=\hbar v/g$, which is the penetration depth of the evanescent states in the photoinduced gap.

We consider now in more detail the saturation regime of the pumped current which arises when the irradiated region is long compared to the characteristic length of the evanescent states in the gap $L \gg \xi$ (Fig. \ref{fig:pumped_L}). In Eq. (\ref{P}), for $0<\varepsilon<g$, one has $\sin(\sqrt{\varepsilon^2-g^2}L)=\sinh(\sqrt{g^2-\varepsilon^2}L)=\sinh(\kappa L)$. Therefore, for $\kappa L\gg1$, $P(\varepsilon) \rightarrow 0$ in the energy window $\epsilon \in [0,g[$, and the pumped current at the Dirac point reads :
\begin{align}
I_{\mu=0}&=\frac{e\omega}{2\pi} & \quad &\text{for} \quad \hbar\omega<2g, \\
&=\frac{e\omega}{2\pi}-2\frac{e}{h} \int_{g}^{\hbar\omega/2}d\varepsilon \, P(\varepsilon)& \quad &\text{for} \quad \hbar\omega>2g.
\end{align}
In conclusion, at low frequency ($\hbar \omega < 2g$ ) the saturated photocurrent reaches a $L$-independent maximum value $e \omega/(2 \pi)$ when $L$ exceeds the characteristic length $\xi=\hbar v/g$. At high frequency ($\hbar \omega > 2g$ ), the saturation exhibits small Fabry-P\'erot like oscillations around an average plateaus value which increases with depends on the ratio $\hbar \omega/g$ (Fig. \ref{fig:pumped_L}). This average saturated current increases with $\hbar \omega/g$ and it is always limited by the maximal value of the photocurrent $\pi eg/h$, which is reached at very high frequency $\hbar \omega \gg g $ and long irradiated length $L \gg \xi$ (see Eq. \ref{eq:universal} below).

\subsubsection{Frequency dependence of the pumped current}

For $\hbar\omega<2g$, the second term in Eq. (\ref{eq:pumped}) vanishes which means that the helical edge state pumps exactly one electron per cycle of the electromagnetic drive.

\begin{figure}[h]
	\begin{center}
		\includegraphics[width=8cm]{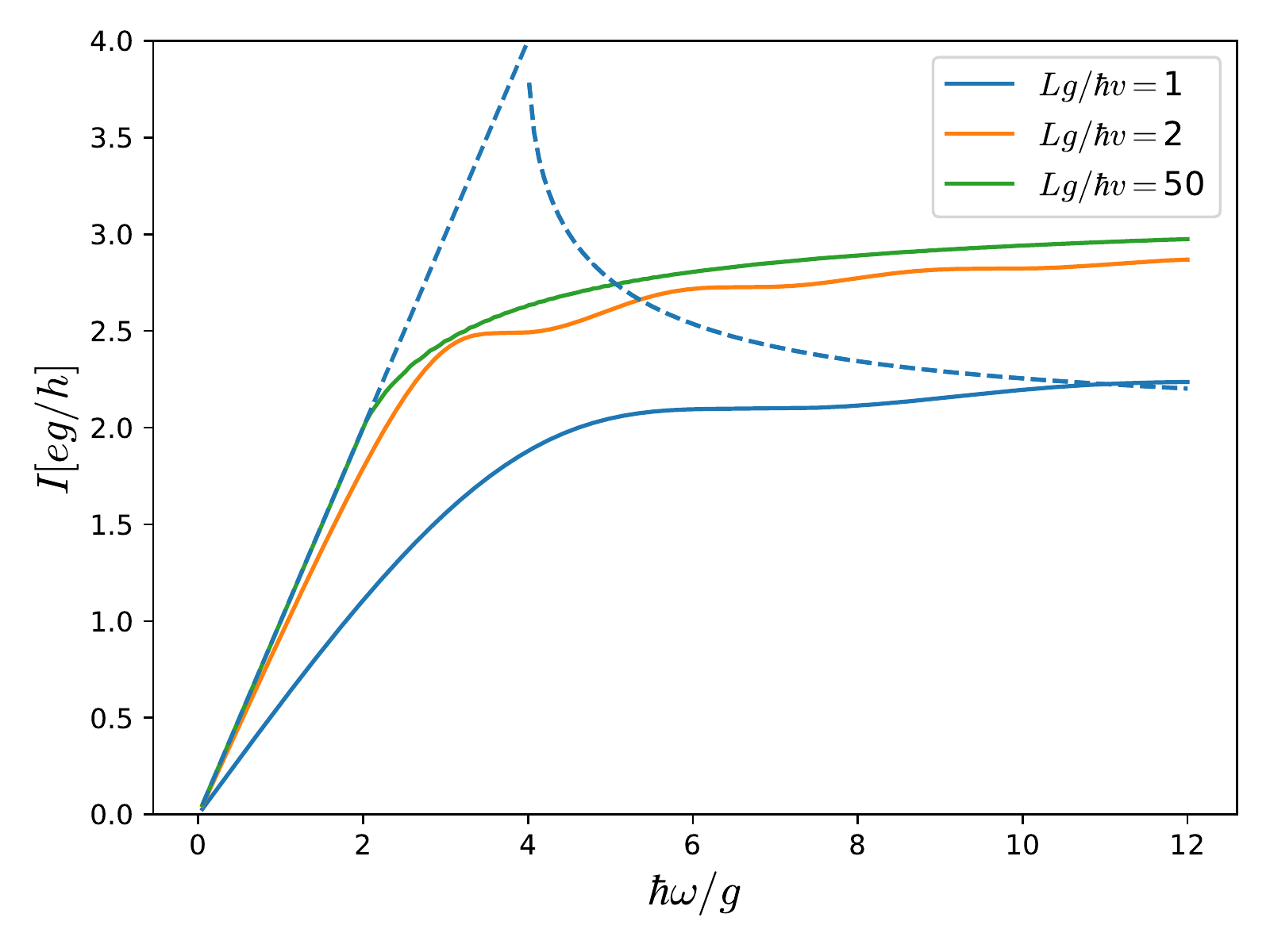}
		\caption{Pumped current as a function of the driving frequency for various lengths of the irradiated region (solid curves), compared with Dora et al. \cite{Dora2012} (dashed curve). For a long ribbon and for frequencies $\hbar\omega<2g$, the current is proportional to $\omega$, which corresponds to the adiabatic quantized charge pumping. For $\hbar\omega>2g$, the current increases and reaches $\pi\frac{eg}{h}$ as $\omega\rightarrow\infty$.}
		\label{fig:pumped_current}
	\end{center}
\end{figure}

For $\hbar\omega>2g$, we can rewrite the current as :
\begin{align}
I_{\mu=0}&=2\frac{eg}{h}+2\frac{e}{h}\int_{g}^{\hbar\omega/2}d\varepsilon \, (1-P(\varepsilon)) \, , \label{eq:current_longL}
\end{align}
where the second term has a finite $\omega \rightarrow \infty$ limit because $1-P(\varepsilon)$ tends to $0$ as $\omega$ tends to infinity (Fig. \ref{fig:trans_coeffs}).

Finally, Fig. \ref{fig:pumped_current} shows the pumped current as a function of the frequency of the driving for various ribbon lengths. As the length of the sample is increased above the characteristic length of the evanescent state $\xi=\hbar v/g$, the curve tends to the limit given by the green curve corresponding to a long ribbon ($L=50\xi$). In this limit, we can clearly see the separation between the low-frequency ($\hbar\omega<2g$) regime and the high-frequency ($\hbar\omega>2g$) regime. We compare our results with Dora \textit{et al.} \cite{Dora2012} where the current is calculated for the infinite system. We obtain the same result at low frequencies for the adiabatic charge pumping regime where the current is linearly proportional to the frequency. Dora \textit{et al.} considered the filling of the bands according to the average energy $\bar{E}_\alpha=\langle\langle u_\alpha| H(t)|u_\alpha\rangle\rangle$, which is a different prescription than the Floquet-Landauer-B\"uttiker formalism used here. At high-frequency, we find the same limiting behaviour except for the presence of an interference term originating from the leads, and a rounding of the cusp of the photocurrent separating the low and high frequency regimes. This rounding and the overall curve for the phorocurrent is very similar to Vajna \textit{et al.} \cite{Vajna2016}, where dissipation is taken into account by coupling (explicitely) the edge state to a bosonic bath. In our model, dissipation occurs (implicitely) in the leads, namely in the external fermionic baths.

The photocurrent, plotted in Fig. \ref{fig:pumped_current}, actually reaches a plateau with increasing $\omega$. Performing the integral in Eq. (\ref{eq:current_longL}) in the case $\hbar\omega/g\rightarrow\infty$ and $Lg/\hbar v\gg1$ gives the value of this high-frequency maximal photocurrent for a long edge :
\begin{equation}
\lim_{\hbar\omega/g,L\rightarrow\infty}I_{\mu=0}=\pi\frac{eg}{h}
\label{eq:universal}
\end{equation}

\subsubsection{Pumped current for non-zero doping :}

\begin{figure}[h]
	\begin{center}
		\includegraphics[width=8cm]{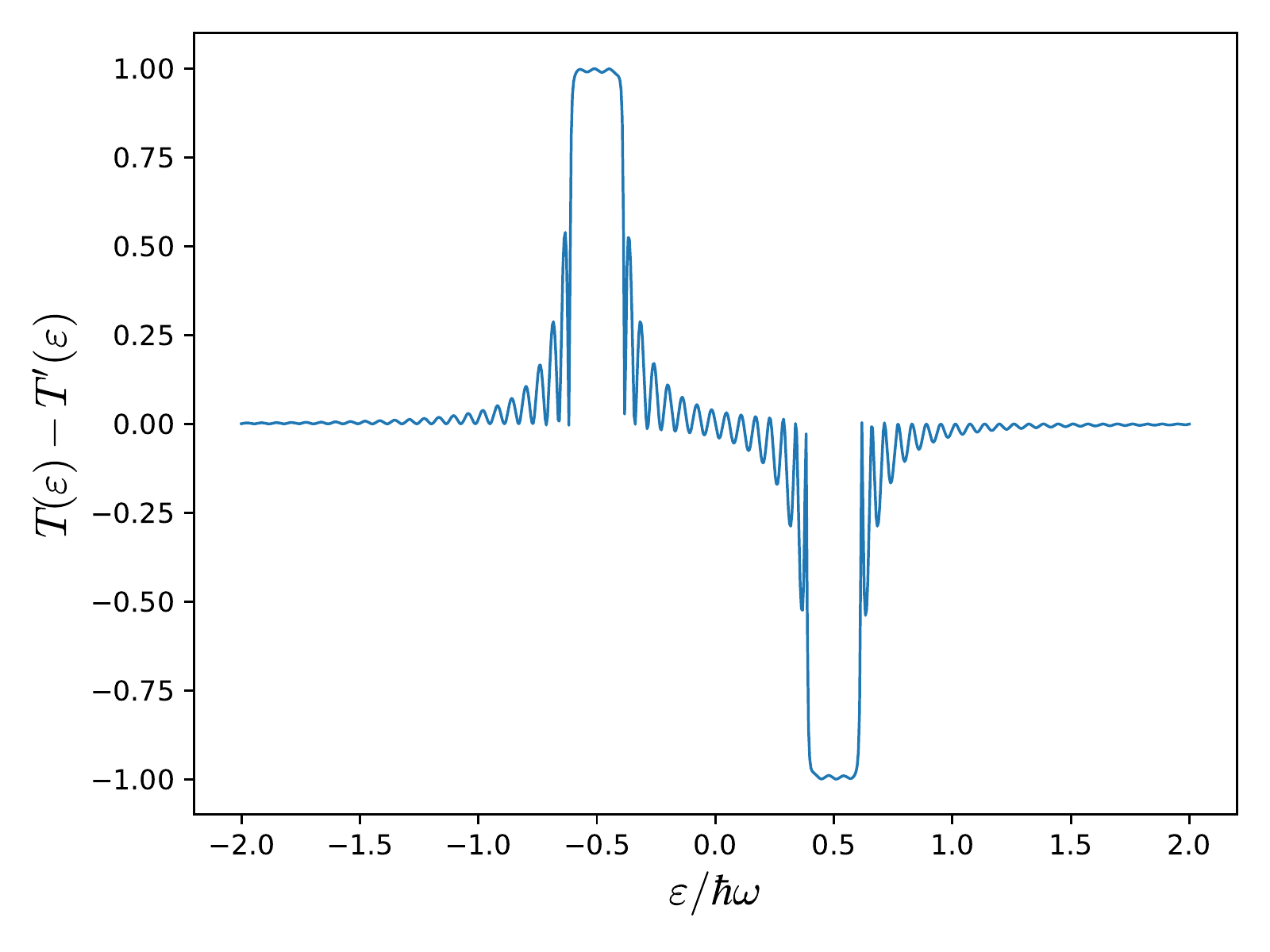}
		\caption{Net transmission probability $T(\varepsilon)-T'(\varepsilon)$ at high-frequency ($g=0.1\hbar\omega$) and for $L=5\hbar v/g$. The current is obtained by integrating from $\varepsilon=-\Lambda$ to $\varepsilon=\mu$.}
		\label{fig:transL-R}
	\end{center}
\end{figure}
In the case of a non-zero doping, we use Eq. (\ref{eq:pumped_diff}) or (\ref{eq:pumped_current}) to calculate the current. Fig. \ref{fig:transL-R} shows the net transmission probabilities $T(\varepsilon)-T'(\varepsilon)$ to be integrated from $-\Lambda$ to $\mu$ to obtain the pumped current. The net transmission equals $1$ around $\varepsilon=-\hbar\omega/2$, while it is equal to $-1$ close to $\varepsilon=\hbar\omega/2$. These peaks have width $2g$. When the chemical potential is set to zero, only the states at quasi-energy smaller than $\mu=0$ contribute to the current which leads to a positive current along the $x$ direction. However, when the chemical potential is increased above $\mu=\hbar\omega/2$, the current originating from states at $\mu=-\hbar\omega/2$ and $\mu=\hbar\omega/2$ will cancel each other, leading to a vanishing net current.

\begin{figure}[t!]
	\begin{center}
		\includegraphics[width=8cm]{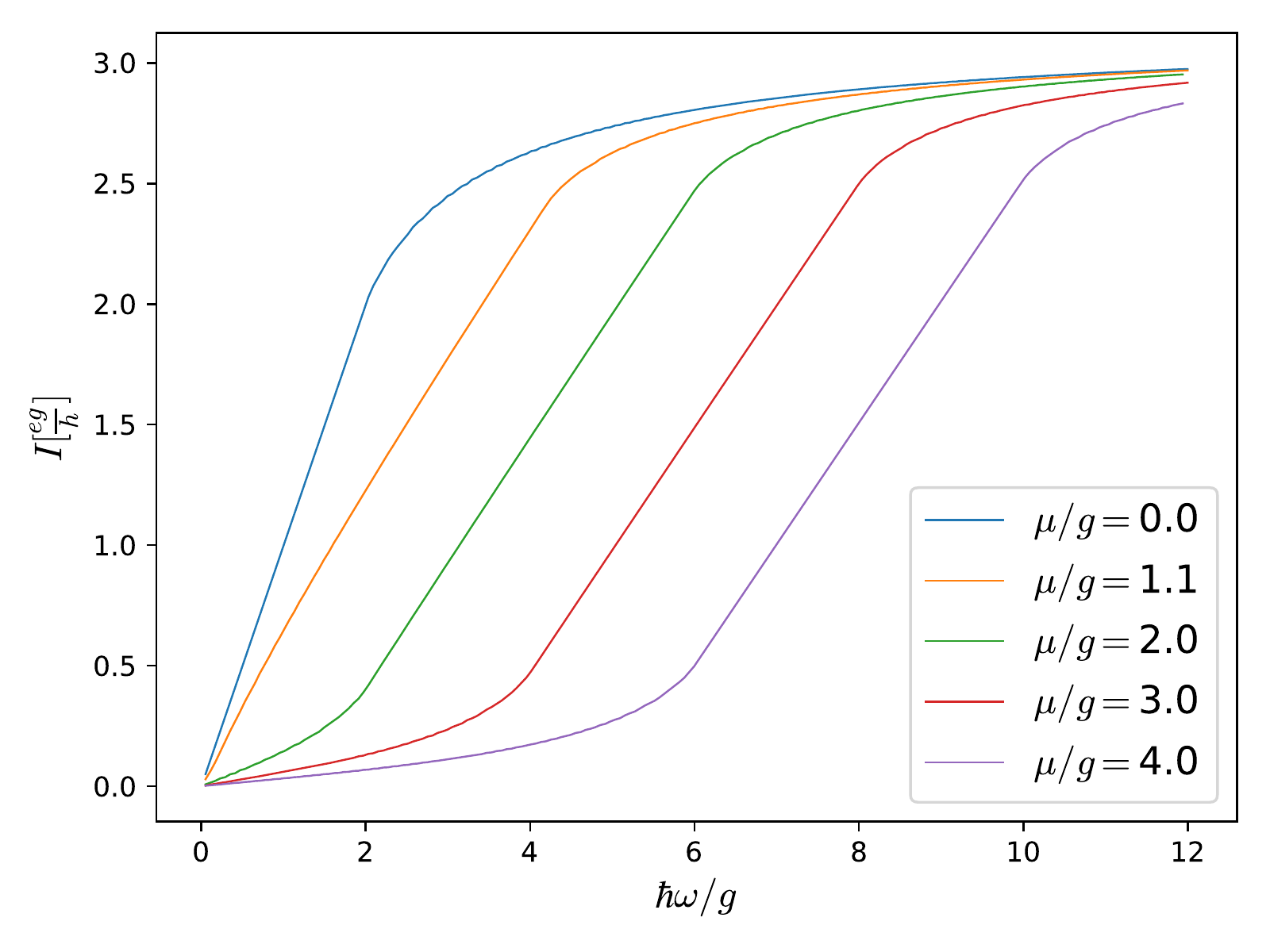}
		\caption{Pumped current as a function of the frequency for various values of the doping $\mu$ in the irradiated region and for $L=50\hbar v/g$. At low frequencies, the effect of the doping is to reduce the conductance.}
		\label{fig:pumped_current_mu}
	\end{center}
\end{figure}

Fig. \ref{fig:pumped_current_mu} shows the pumped current as a function of the frequency for various values of the chemical potential $\mu$. The effect of the chemical potential is to reduce the current in the low frequency regime. In fact, the current will be close to zero when the frequency is small such that $\hbar\omega/2+g<\mu$ because in that case both peaks of Fig. \ref{fig:transL-R} are integrated and the current vanishes. However, in the high frequency regime $\hbar\omega\gg g$ and $\hbar\omega/2>\mu$, the pumped current still reaches the value given by Eq. (\ref{eq:universal}).

We recover a linear behavior as a function of $\omega$ when the frequency is in the range $\hbar\omega\in[2(\mu-g),2(\mu+g)]$. This range corresponds to a scenario where the chemical potential is in the dip located around $\hbar\omega/2$ in Fig. \ref{fig:transL-R}. Finally, at high frequency, when the chemical potential $\mu$ is small compared to $\hbar\omega$, we recover the zero-doping scenario where only the peak at quasi-energy $\varepsilon=-\hbar\omega/2$ contributes to the current. In that case, independently of the doping, the pumped current reaches the value of the undoped case $I=\pi eg/h$.

\subsection{Photoconductance}

\begin{figure}[t]
	\begin{center}
		a)\includegraphics[width=8cm]{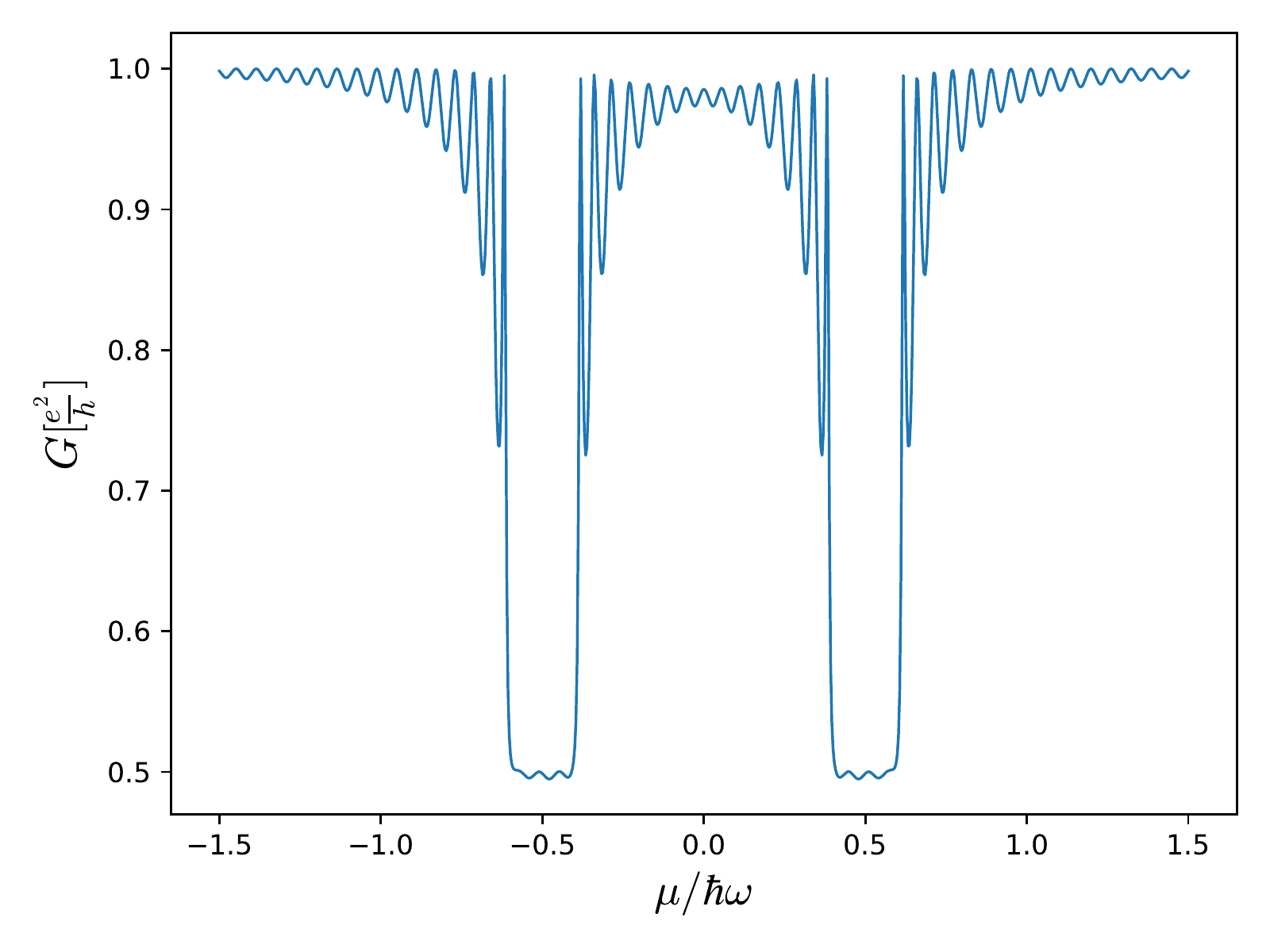}
		b)\includegraphics[width=8cm]{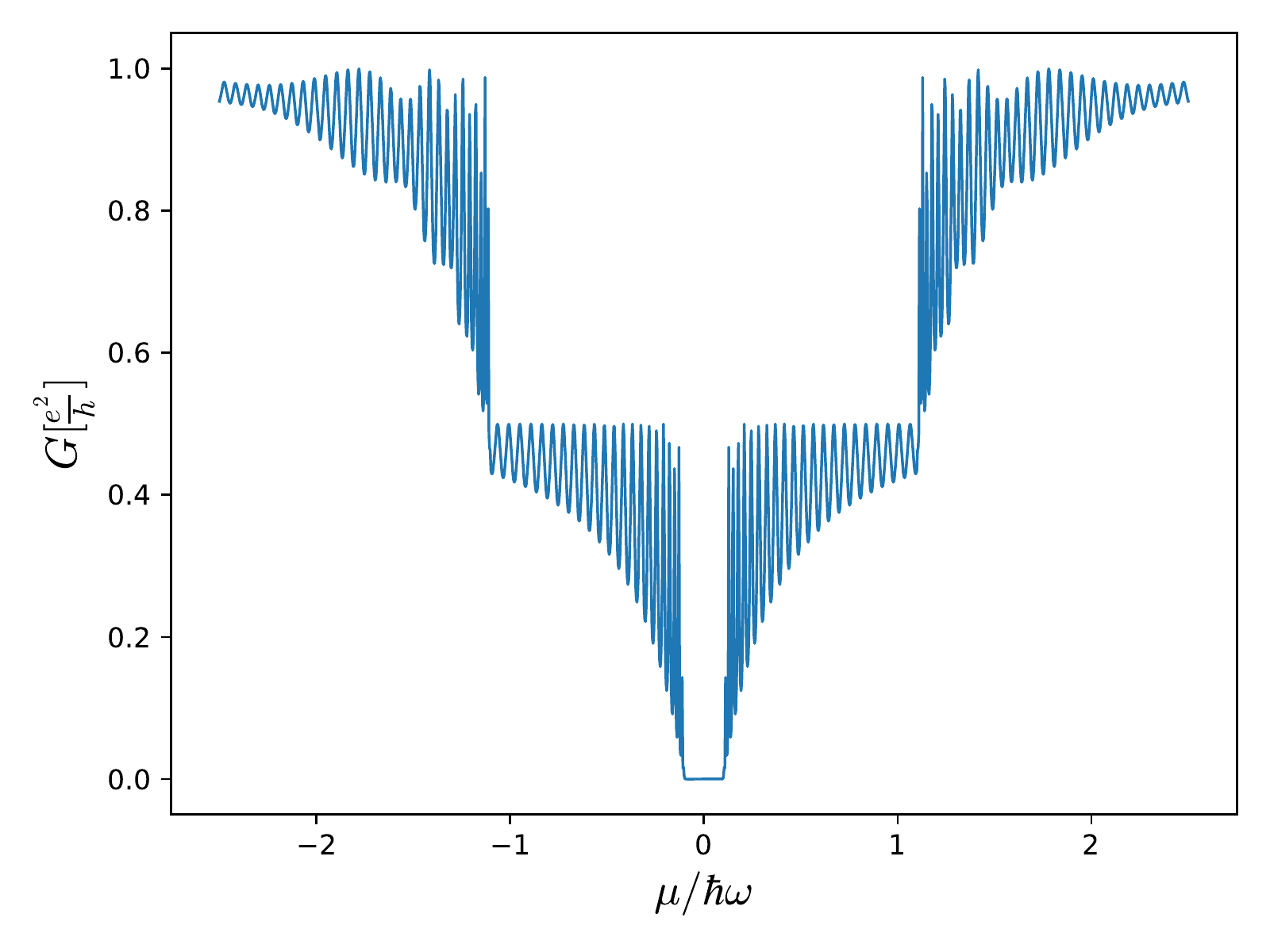}
		\caption{Conductance of the irradiated edge state as a function of the chemical potential for a driving strength a) $g=0.1\hbar\omega$ and $L=5\hbar v/g$ corresponding to the high-frequency regime and b) $g=0.6\hbar\omega$ and $L=30\hbar v/g$ corresponding to the low-frequency regime. Here, the plotted conductance is the one carried by the irradiated edge only (upper edge on Fig. 1). The non irradiated edge also contributes by a trivial quantum of conductance $e^2/h$ which adds up to the plotted conductance to form the total 2-terminal conductance corresponding to Fig.1.}
		\label{fig:conductance}
	\end{center}
\end{figure}

In this section, we investigate the effect of a voltage bias $V$ between the leads. Starting from Eq. (\ref{eq:current}), the current can be expressed as :
\begin{align}
I&=\frac{e\omega}{2\pi}- \frac{e}{h}\int_{\mu_L-\hbar\omega/2}^{\mu_R+\hbar \omega/2}d\varepsilon \,  P(\varepsilon).
\end{align}
We consider a chemical potential $\mu_L=\mu+eV/2$ in the left lead and $\mu_L=\mu-eV/2$ in the left lead which allows us to write the current $I(V)$ as :
\begin{align}
I(V)&=
I(0)+\frac{e}{h}\int_{\mu-\hbar\omega/2}^{\mu-\hbar \omega/2+eV/2}d\varepsilon P(\varepsilon) \\ &\qquad\qquad+\frac{e}{h}\int_{\mu+\hbar\omega/2-eV/2}^{\mu+\hbar\omega/2}d\varepsilon P(\varepsilon),\nonumber\\
&=I(0)+\frac{e^2 V}{2h}\left[P \left(\mu-\frac{\hbar\omega}{2} \right)+P\left(\mu+\frac{\hbar\omega}{2} \right)\right]
\end{align}
where $I(0)=I(V=0)$ is the usual current in absence of bias given by Eq. (\ref{eq:current}). We have considered a small bias such that the function $f(\varepsilon)$ varies weakly with $V$. Finally, the expression for the conductance reads :
\begin{equation}
    G=\frac{\partial I}{\partial V}=\frac{e^2}{2h}\left[T(\mu)+T'(\mu)\right] \, .
\end{equation}
Fig. \ref{fig:conductance} shows the conductance of the irradiated egde state as a function of the chemical potential. In Fig. \ref{fig:conductance}.a), which corresponds to the high-frequency regime, we observe two dips in the conductance located at $\mu=-\hbar\omega/2$ and $\mu=\hbar\omega/2$. These dips have a chemical potential width $2g$ and correspond to the gaps in the quasi-energy spectrum (Fig. \ref{fig:spectrum}.a)). Inside these dips, the conductance doesn't vanish and remains close to $G=0.5 \, e^2/h$ due to the presence of propagating state belonging to the other replicas. Away from these dips, the conductance is close to $e^2/h$, corresponding to perfect transmission, and shows oscillations characteristic of Fabry-P\'erot interferences. We have evaluated and discussed here the conductance carried by a single edge (the irradiated one). In the experimental set-up shown in Fig. \ref{fig:geometry}, the lower non-irradiated edge also contributes to the conductance up to a single quantum of conductance $e^2/h$ for any value of the chemical potential.

Fig. \ref{fig:conductance}.b) shows the conductance in the low-frequency regime $g=0.6 \, \hbar\omega$ corresponding to the spectrum of Fig. \ref{fig:spectrum}.b). In this regime, we observe the central quasi-energy gap at $\varepsilon=0$ where no propagating states originating from the replicas $n=0$ and $n=1$ exist. We also observe a plateau in the range $\mu/\hbar\omega\in[-1,1]$ where the conductance oscillates but is bounded below $G=0.5 \, e^2/h$. In this range, the current is carried by the bands $(\alpha,n)=(+,0)$ for a positive chemical potential and $(-,1)$ for a negative chemical potential corresponding to the bands in green and orange respectively on Fig. \ref{fig:spectrum}.b). As the chemical potential is increased such that $|\mu|>1$, another band is accessible and the conductance is bounded below $G=e^2/h$.

In conclusion, the Floquet gap structure of the irradiated edge state can be scanned by measuring the differential conductance. In this transport setting, only the two quasi-energy bands $n=0$ and $n=1$ contribute to the current. It is also possible to discriminate between the two frequency regimes : in the high-frequency regime ($\hbar\omega>2g$), the two small dips are predicted in the conductance at quasi-energy $\varepsilon=\pm\hbar\omega/2$ while two "nested" dips centered at $\varepsilon=0$ are predicted in the low-frequency regime ($\hbar\omega<2g$).

\subsection{Experimental realization}

We discuss here the regime of the laser parameters for which these two charge pumping regimes can be observed experimentally. The first constraint is that the laser frequency must be smaller than the bulk band gap of the material. Recent realizations of the QSH effect in WTe$_2$ crystal present gaps of the order of $50-100$ meV\cite{Tang2017,Wu2018}, which puts an upper bound on the laser frequency of the order of $10$ THz. For a laser power of 100 mW/$\mu m^2$, the typical electric field strength is $E_0=6\times10^6\text{V/m}$. The Zeeman coupling constant is thus of the order $g=g_\text{eff}\mu_BB_0=g_\text{eff}\mu_BE_0/c\approx10^{-5}\text{eV}$, where the effective $g$-factor which can be enhanced ($g_\text{eff}\approx20-50$) in materials with strong spin-orbit coupling like HgTe/CdTe \cite{Dora2012}. Such a Zeeman interaction strength gives rise to a characteristic length of the evanescent states of the order of 1 $\mu m$. For a laser frequency of 1 THz, we are in the weak coupling $g\ll\hbar\omega$ (high-frequency) regime, and the pumped current is thus of the order of $I=\pi eg/h=1 \,$ nA.

\section{Conclusion}

 In this paper, we have studied the electronic transport properties of a single irradiated helical edge state of a QSH insulator, extending previous works\cite{Dora2012,Vajna2016} by investigating the effect of a finite length between the leads. We provide an analytical expression (Eqs. \ref{P},\ref{eq:pumped}) allowing to cover the complete crossover from $L=0$ (no irradiation, no current) to very long $L$ (saturated photocurrent) as shown in Fig. \ref{fig:pumped_L}. When the chemical potential of the edge state is located at the Dirac point (band crossing), a pumped photocurrent is predicted in the absence of bias between the leads. In the low-frequency regime, this current has the same behaviour as predicted in Dora \textit{et al.}\cite{Dora2012} which corresponds to a quantized pumped charge per unit cycle. However, in the high-frequency regime, the effect of the leads is to round off the crossover between the quantized and unquantized regimes (Fig. \ref{fig:pumped_current}), in comparison to the calculation in the infinite system without dissipation. Interestingly, a similar behavior of the photocurrent has been predicted in \cite{Vajna2016} using a rather different model : an infinite helical liquid coupled to an external bath. Using an external gate, the chemical potential can be tuned, which tends to to reduce the pumped current in the low-frequency regime. Finally, we have also investigated the effect of a voltage bias, and computed the corresponding photoconductance of the edge state. We found that differential conductance is a good tool to explore the quasi-energy Floquet spectrum of the edge state.

\label{sec:conclusion}

\bibliographystyle{apsrev4-1}
\bibliography{bibl}
\end{document}